# A modified model of a single rock joint's shear behavior in limestone specimens

Dindarloo Saeid R [a*], Siami-Irdemoosa Elnaz [b]

[a] Department of Mining & Nuclear Engineering, Missouri University of Science &Technology, Rolla, MO, USA
[b] Department of Geosciences & Geological & Petroleum Engineering, Missouri University of Science &Technology, Rolla, MO, USA

ABSTRACT

The shear behavior of a single rock joint in limestone specimens, under a constant normal load (CNL), was analyzed in this study. Test specimens with different asperity roughness were prepared and tested. Goodman's model of a rock joint's shear behavior, under CNL, was modified to render a better representation of the data obtained. The model's applicability was validated. The proposed model showed better correlation with experimental data. It also, requires fewer variables. The steps to calculate all the necessary variables for the model are discussed.

**Nomenclature**

$\tau$ = Shear Stress

$\tau_p$ = Peak Shear Strength

$\tau_r$ = Residual Shear Strength

u = Shear Displacement

$u_p$ = Shear Displacement at Peak Shear Strength

$u_r$ = Residual Displacement

$\sigma_n$ = Normal Stress

$\sigma_T$ = Transitional stress in Ladanyi _ Archambault

$\varphi$ = Internal Friction Angle

$a_s$ = Proportion of Total Joint Area Sheared Through Asperities

$s_r$ = Shear strength of asperities

$k_1, k_2$ = Empirical Constants in Ladanyi Archambault

$C_1, C$ = First and Second Constants of the Experiment in the Proposed Model

$\dot{v}$ = Secant Rate of Dilatancy at Peak Shear Strength

i= Arctan($\dot{v}$)

$i_m$ = Triangular Asperity Angle

Z = Shear Strength Factor in the Proposed Model

----------------
* Corresponding author: Email: srd5zb@mst.edu, Tel: (+1) 573- 201-0737

## 1. Introduction

Rock joints are mechanical discontinuities that have geological origins. In general, the strength and deformability properties of these discontinuities are quite different from those of intact rock. In many cases, the discontinuities completely dominate both the shear and the deformation behavior of the in situ rock mass in given stress conditions [1-2]. Engineers in the mining, civil, and petroleum industries often face problems that are associated with jointed rock masses. Rock joint's shear behavior must be examined comprehensively to understand the jointed rock mass mechanical behavior. Many applications could benefit from the study of joints at a smaller scale, such as petroleum and energy recovery applications [3]. A number of researchers have tried to model the shear behavior of a single rock joint under laboratory conditions-most use the direct shear test. The test is conducted under two major boundary conditions. A direct shear test under constant normal load (CNL) and a direct shear test under constant normal stiffness (CNS). A CNL is used when the rock can dilate freely i.e. with constant normal load under shear displacement. This situation is typically encountered in surface rock structures such as rock slopes. In case, the joint is constrained with surroundings materials and cannot dilate freely upon shearing, the normal load will increase. This load's curve is controlled by the stiffness of surrounding rocks. The CNS condition is typically encountered in deep underground cavitations. The shear behavior of rock joints is not simply controlled by boundary conditions (i.e., either CNL or CNS). It is also controlled by a number of other important factors, including the intact rock properties, joint roughness, shear rate, and filling materials [4-5].

A comprehensive mathematical model that considers all of these effective variables has not been developed. The application of experimental methods and models is necessary to addressing the difficulties of modeling this complex behavior analytically [6]. Experimental results are useful both in modeling and calibrating several of the model's parameters. They are also useful in validating the results. Direct shear tests under the CNL condition were conducted on natural rock joints in this study. The results were used to render an experimental equation for the shear behavior. Tests specifications, specimens, and materials are introduced in Sec. 2. The Goodman's model under the CNL condition and the proposed model are discussed in Secs. 3-4. Finally, Sec. 5 concludes the paper.

## 2. Specimens and tests specifications

Fourteen limestone specimens were collected and prepared for the purpose of understanding the shear behavior of joints in limestone rocks. These specimens were collected from a dam site located inside a limestone zone. The direct shear test procedure conducted by Bandis et al [7] was used in this study. The material's basic properties were examined through a series of direct shear tests on solid blocks. The basic sliding resistance tests were performed on planar solid surface under various normal stresses. The shear displacement rate was 0.5 mm/min. Triangular asperities with different angles (from 4 to 20 degrees) were presented in the samples. These asperity angles have considerable effect on the shear's behavior [8].

## 3. Model Description:

The three-line model proposed by Goodman [9] served as the mathematical model's starting point. Equations (1-3) define the shear stress versus shear displacement for the region before peak shear strength, between peak and residual shear strengths, and after residual, respectively.

$$\tau = \frac{\tau_p}{u_p} u , \qquad u < u_p \qquad (1)$$

$$\tau = \left(\frac{\tau_p - \tau_r}{u_p - u_r}\right) u + \left(\frac{\tau_r u_p - \tau_p u_r}{u_p - u_r}\right), \quad u_p \leq u \leq u_r \qquad (2)$$

$$\tau = \tau_r , \qquad u > u_r \qquad (3)$$

These equations are plotted in Fig. 1.

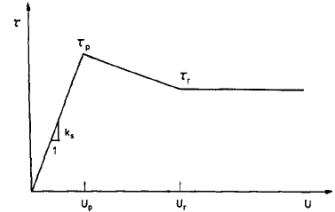

Fig. 1 Goodman's model for the shear behavior of rock joints.

In this study, a two-domain model is proposed. The first part is the same as Goodman's, as it predicts the joint behavior almost exactly in the same way as the experimental results. However, the second and third parts of Goodman's are simplified mathematical representation of the actual behavior. These deviated from the actual test results in this case. In this paper, the after region peak is modeled as a non-linear functional relationship with a shear displacement in form of $\tau \propto \frac{1}{u}$ as calculated in equation (5). The curve in Fig. 2 has a better correlation than Fig. 1 with experimental results. The proposed model is defined as below:

$$\tau = \frac{\tau_p}{u_p} u , \qquad 0 \leq u \leq u_p \qquad (4)$$

$$\tau = \tau_r + (\tau_p - \tau_r)\frac{u_p}{u} , \qquad u > u_p \qquad (5)$$

Only three parameters are required for the complete mathematical modeling of a single joint's shear behavior under the CNL condition, as verified in (4) and (5). These parameters include the peak shear strength, the residual shear strength, and shear displacement at the peak shear strength. Hence, these three parameters should be defined from the intact rock properties, the joint geometry, the mechanical properties, loading type, and the loading rate. Saeb and Amadei [10] applied (6) to estimate the peak shear strength.

$$\tau_p = \sigma_n \tan(\varphi + i)(1 - a_s) + a_s s_r \qquad (6)$$

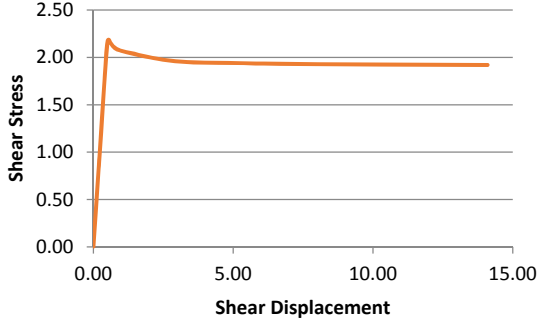

Fig. 2 Proposed model for the shear behavior of a single rock joint.

Where $a_s$ and $S_r$ are the proportion of the total joint area sheared through asperities and the shear strength of asperities, respectively. An accurate measurement of these parameters is not practical, particularly in in-situ tests. Ladanyi and Archambault [11] proposed that the following formula be used to calculate as and i. This requires determination of further unknowns [10] as shown in (7-9).

$$a = 1 - \left(1 - \frac{\sigma_n}{\sigma_T}\right)^{k_1} \quad (7)$$

$$i = \arctan(\dot{v}) \quad (8)$$

$$\dot{v} = \left(1 - \frac{\sigma_n}{\sigma_T}\right)^{k_2} \cdot \tan(i_0) \quad (9)$$

Where $k_1$ and $k_2$ are empirical constants and $\sigma_T$ is a transitional stress. The uniaxial compressive strength of the intact rock can be used as an estimate of $\sigma_T$ [11].

More unknowns need to be defined before $\tau_p$ in (6) can be calculated. Several of these parameters can be measured accurately; several are only estimations. Thus, the application of (6) does not guarantee exact results that are comparable to actual ones obtained either by laboratory or in-situ tests. Equation (10) was used in this study to obtain a sufficiently accurate estimate of the peak shear strength while using minimum number of variables. The values obtained from (10) were compared with the actual values obtained and plotted in Fig. 3. This shows a very good agreement between the two data sets.

$$\tau_p = \sigma_n \tan(\varphi + i_m) \quad (10)$$

Goodman proposed the following model for $\tau_r$ at different normal stresses [9].

$$\tau_r = \tau_p \left(B_0 + \frac{1 - B_0}{\sigma_T} \sigma_n\right), \quad \text{for } \sigma_n < \sigma_T \quad (11)$$

Where $B_0$ is the ratio of the residual strength to the peak shear strength at a zero normal stress and equation (12) holds for the residual strength.

$$\tau_r = \tau_p, \quad \text{for } \sigma_n \geq \sigma_T \quad (12)$$

The same difficulty arises when (11) and (12) are applied; because $\tau_r$ is defined as a function of $\tau_p$. Furthermore, one more parameter ($B_0$) is added. Equations (13) and (14) were used in this study; because they require a minimal amount of variables to achieve good approximates for the residual shear strength. Again, the values obtained from (13) were compared with actual values obtained for $\tau_r$. This comparison is illustrated in Fig 4. The results taken from 14 random samples exhibited an acceptable agreement between the two data sets.

$$\tau_r = \sigma_n \tan(\varphi), \quad \sigma_n < \sigma_T \quad (13)$$

$$\tau_r = \tau_p, \quad \sigma_n \geq \sigma_T \quad (14)$$

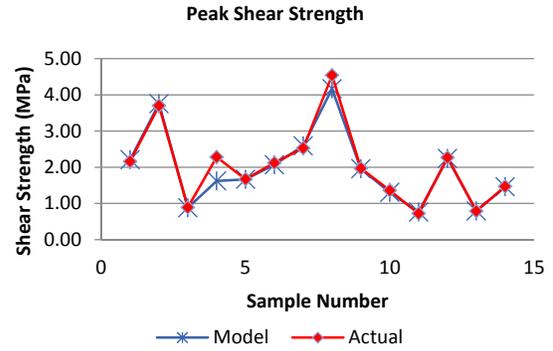

Fig. 3 Comparison between the peak shear strength calculated in (10) and the experimental data.

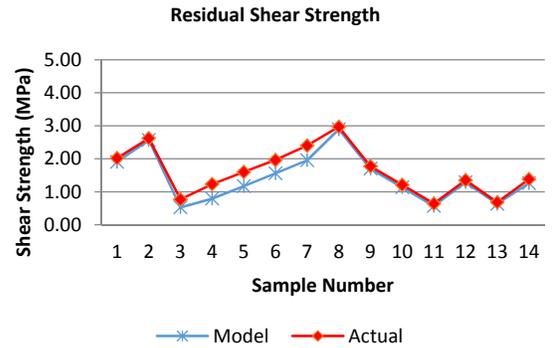

Fig. 4 A comparison between the residual shear strength calculated in (13) and the experimental data.

The only remaining variable from (4) and (5) is the shear displacement at the peak shear strength ($U_p$). The following empirical formula was proposed for purpose of calculating up:

$$u_p = C_1 e^{C_2 Z} \qquad (15)$$

Where $C_1$ and $C_2$ are the experiment's first and second constants, respectively. These constants depend on joint roughness and the rate of the shear test (based on result of sensitivity analysis in this study) and will be defined experimentally. The $Z$ is a joint strength factor that is dependent on $\tau_p, \tau_r$, and $\sigma_n$. It is calculated in (16).

$$Z = \sqrt{\tau_p \cdot \tau_r} + \sigma_n \qquad (16)$$

Figure 5 is an illustration of the shear displacement ($U_p$) versus the introduced strength factor ($Z$). The real observed data for both the peak and the residual shear stresses, at different applied normal stresses, for 14 samples was used to depict this relationship. An exponential relationship is a good approximate of the relationship as suggested in (15). The $C_1$ and $C_2$ constants are 0.25 and 0.19 for this study, respectively.

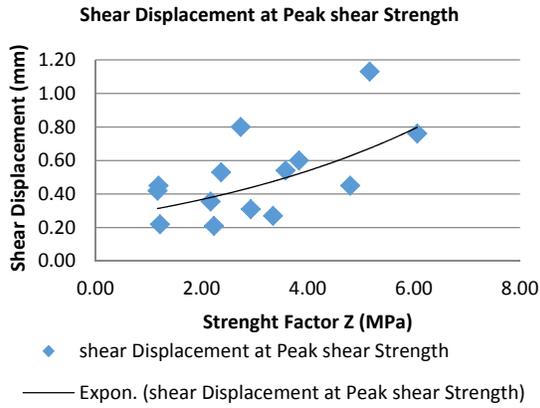

Fig. 5 Illustration of equation 15 based on estimated model's values.

The data plotted in Fig. 6 illustrates that a good correlation exists between the $U_p$ proposed by (15) and the actual values obtained from direct shear tests.

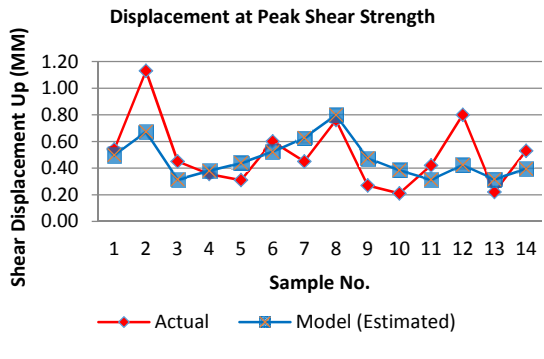

Fig. 6 Actual versus estimated $U_p$.

In conclusion, a good approximation for parameters $\tau_p, \tau_r$ and up, is given in Figs. 3, 4, and 6 respectively. Having defined the three parameters $\tau_p, \tau_r$ and $U_p$ from (10), (13), and (15); equations (4) and (5) can predict a rock joint's shear behavior under CNL condition. The applicability of the new model will be compared with the actual data from direct shear tests.

## 4. Model Validation

Results of direct shear tests and the proposed model are compared in the following section. Comparisons between three randomly selected samples and the actual results are shown in Figures 7-9 and Table 1.

*4.1. Model Equations*

The procedure of equations derivation is elaborated for the first sample in the following. The results gathered from the remaining two experiments are illustrated in Figures 8 and 9. They are summarized in Table 1.

From Equation (10), substituting, $\sigma_n, \varphi,$ and $i_m$ (This data is available from testing the specimen's material, asperities, and CNL machine):

$$\tau_p = \sigma_n \tan(\varphi + i_m) \qquad (17)$$

$$\tau_p = 1.531 \tan(51.24 + 4) = 2.21 \, (MPa) \qquad (18)$$

From Equation (13), substituting, $\sigma_n$:

$$\tau_r = \sigma_n \tan(\varphi) \qquad (19)$$

$$\tau_r = 1.531 \tan(51.24) = 1.91 \, (MPa) \qquad (20)$$

From Equation (16) for $Z$:

$$Z = \sqrt{\tau_p \cdot \tau_r} + \sigma_n \qquad (21)$$

$$Z = \sqrt{2.21 \times 1.91} + 1.531 = 3.586 \, (MPa) \qquad (22)$$

From Equation (15), substituting, $C_1$ and $C_2$ with 0.2508 and 0.191, respectively,

$$u_p = C_1 e^{C_2 Z} \qquad (23)$$

$$u_p = 0.25 e^{0.19(3.586)} = 0.50 \qquad (24)$$

From Equations (4) and (5), substituting, equations (17), (20), and (24)

$$\tau = \frac{\tau_p}{u_p} u \, , \qquad 0 \le u \le u_p \qquad (25)$$

## Table 1

Test specifications and model estimations

| Sample | $i_m$ (°) | $\varphi$ (°) | Actual $u_p$ (mm) | Actual $\tau_p$ (MPa) | Actual $\tau_r$ (MPa) | $\sigma_n$ (MPa) | Actual $Z$ (MPa) | Estimated $u_p$ (mm) | Estimated $\tau_p$ (MPa) | Estimated $\tau_r$ (MPa) | Estimated $Z$ (MPa) |
|---|---|---|---|---|---|---|---|---|---|---|---|
| 1 | 4 | 51.24 | 0.54 | 2.16 | 2.02 | 1.53 | 3.62 | .50 | 2.21 | 1.91 | 3.58 |
| 2 | 10 | 51.24 | 1.13 | 3.70 | 2.62 | 2.06 | 5.17 | .67 | 3.76 | 2.57 | 5.17 |
| 3 | 10 | 48.15 | 0.76 | 4.54 | 2.97 | 2.59 | 6.26 | .81 | 4.17 | 2.89 | 6.06 |

$$\tau = \tau_r + (\tau_p - \tau_r)\frac{u_p}{u}, \quad u > u_p \quad (26)$$

$$\tau = \frac{2.21}{0.5}u, \quad u \leq 0.5 \quad (27)$$

$$\tau = 1.91 + (2.21 - 1.91)\frac{0.5}{u}, \quad u > 0.5 \quad (28)$$

Figure 7 is a curve of the shear behavior for sample No. 1.

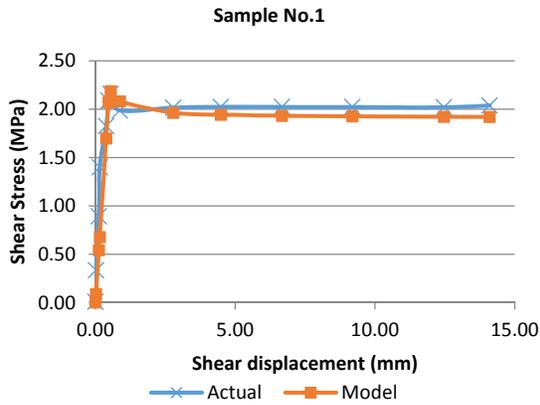

Fig. 7 Comparison between the estimated results and the actual values obtained for the first sample.

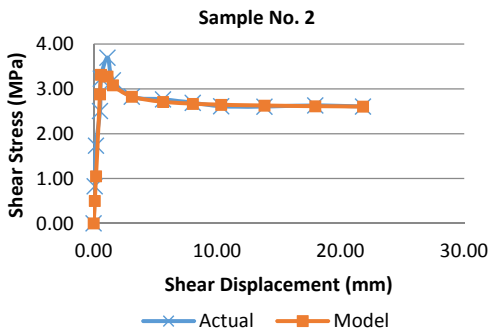

Fig. 8 Comparison of estimated results with actual values for the second sample.

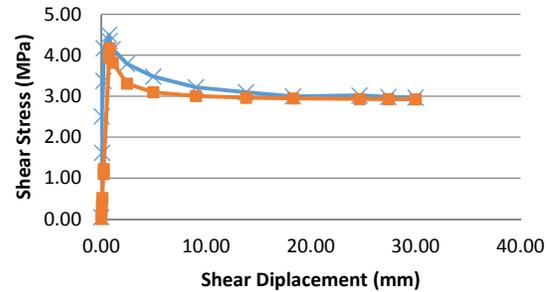

Fig. 9 Comparison between the estimated results and the actual values obtained for the third sample.

## 5. Conclusion

Direct shear tests, conducted under constant normal loads, were used to measure the shear stress versus the shear displacement of fourteen limestone specimens at a fixed shear rate. The obtained shear curves are completely non-linear after the peak shear strength. The conventional three straight-line Goodman's model deviated from the experimental results, particularly in the post-peak strength region. A better mathematical non-linear model was proposed and validated. The development of this model to other rock joints and test specifications is expected.